*Research Article*

# Efficient and Secure Routing Protocol for Wireless Sensor Networks through Optimal Power Control and Optimal Handoff-Based Recovery Mechanism

S. Ganesh[1] and R. Amutha[2]

[1] *Electronics and Communication Engineering, Sathyabama University, Tamil Nadu, Chennai 600075, India*
[2] *ECE Department, SSN College of Engineering, Tamil Nadu, Chennai 603110, India*

Correspondence should be addressed to S. Ganesh, ganesh8461@gmail.com

Advances in wireless sensor network (WSN) technology have provided the availability of small and low-cost sensor with capability of sensing various types of physical and environmental conditions, data processing, and wireless communication. In WSN, the sensor nodes have a limited transmission range, and their processing and storage capabilities as well as their energy resources are also limited. Modified triple umpiring system (MTUS) has already proved its better performance in Wireless Sensor Networks. In this paper, we extended the MTUS by incorporating optimal signal to noise ratio (SNR)-based power control mechanism and optimal handoff-based self-recovery features to form an efficient and secure routing for WSN. Extensive investigation studies using Glomosim-2.03 Simulator show that efficient and secure routing protocol (ESRP) with optimal power control mechanism, and handoff-based self-recovery can significantly reduce the power usage.

## 1. Introduction

Wireless Sensor Network is widely considered as one of the most important technologies for the twenty-first century. The sensing electronics measure ambient conditions related to the environment surrounding the sensor and transform them into an electrical signal. In many WSN applications [1], the deployment of sensor nodes is performed in an ad hoc fashion without careful planning and engineering. In the past few years, an intensive research that addresses the potential of collaboration among sensors in data gathering and processing and in the coordination and management of the sensing activities was conducted. However sensor nodes are constrained in energy supply and bandwidth. Such constraints combine with a typical deployment of large number of sensor nodes that pose many challenges to the design and management of WSNs and necessitate energy awareness at all layers of the networking protocol stack.

At the network layer, it is highly desirable to find methods for energy-efficient route discovery and relaying of data from the sensor nodes to the base stations, so that the lifetime of the network is maximized. Routing in WSN is very challenging due to the inherent characteristics that distinguish these networks from other wireless networks like mobile ad hoc networks or cellular networks. First, due to the relatively large number of sensor nodes, it is not possible to build a global addressing scheme for the deployment of large number of sensor nodes as the overhead of ID maintenance is high. Thus, traditional IP-based protocols may not be applied to WSN. Second, in contrast to typical communication networks, almost all applications of sensor nodes require the flow of sensed data from multiple sources to a particular base station. Third, sensor nodes are tightly constrained in terms of energy, processing and storage capacities. Thus they require careful resource management. Further, in most application scenarios, nodes in WSNs are generally stationary after deployment except for, maybe, a few mobile nodes. Due to such differences, many algorithms like low-energy adaptive cluster hierarchy (LEACH), power-efficient gathering in sensor information systems (PEGASIS), and virtual grid architecture (VGA) have been proposed for the routing problems in WSNs.

Energy conservation is critical in Wireless Sensor Networks. Replacing or recharging batteries are not an option for sensors deployed in hostile environments. Generally communication electronics in the sensor utilizes most energy. Stability is one of the major concerns in advancement of Wireless Sensor Networks (WSNs). A number of applications of WSN require guaranteed sensing, coverage, and connectivity throughout its operational period. Death of the first node might cause instability in the network. Therefore, all of the sensor nodes in the network must be alive to achieve the goal during that period. One of the major obstacles to ensure these phenomena is unbalanced energy consumption rate. Different techniques have already been proposed to improve energy consumption rate such as clustering, efficient routing, and data aggregation.

In [2], Li et al. have investigated the joint power allocation (PA) issue in a class of MIMO relay systems. By using the capacity and the mean-square error (MSE) as optimization criterion, two joint PA optimization problems have been formulated. As the cost functions derived directly from the capacity and the MSE would lead to nonconvex optimization, two modified cost functions corresponding to a convex problem of the source and the relay power weighting coefficients have been developed. The key contribution of the proposed method lies in the discovery of a tight bound for the capacity and the MSE that simplifies the joint source and relay power allocation into a convex problem. A distinct feature of the new method is that the power allocation within the source and that within the relay are jointly optimal for any given power ratio of the two units.

Li et al. and Yang et al. [3, 4] have studied the joint power allocation problem for multicast systems with physical-layer network coding based on the maximization of the achievable rate. To deal with the nonconvex optimization problem, a high-SNR approximation is employed to modify the original cost function in order to obtain a convex minimization problem, where the approximation is shown to be asymptotically optimal at the high-SNR regime. As an alternative, an iterative algorithm has been developed by utilizing the convexity property of the cost function with respect to a part of the whole power coefficients. Considering the low complexity of the physical layer network coding in the multicast system, the lattice-based network coding that uses the proposed joint power allocation schemes has been suggested.

In this paper, we investigate the performance of ESRP in WSN to attain extra routing information through optimal SNR-based power control mechanism and optimal handoff-based self-recovery features to reduce the power usage and decrease the latency of packet delivery. The rest of this paper is organized as follows. In Section 2, the related work is briefly reviewed and discussed. Then we describe our network model, adversary model, and notations used throughout in this paper in Sections 3, 4, and 5. Simulation results are presented in Section 6. We conclude this paper in Section 7.

## 2. Related Work

The task of finding and maintaining routes in WSNs is nontrivial since energy restrictions and sudden changes in node status cause frequent and unpredictable topological changes. Routing protocols can be classified into three categories, namely, proactive (table driven), reactive (on demand), and Hybrid protocols depending on how the source finds a route to the destination. In proactive protocols sensors advertise their routing state to the entire network to maintain a common (partially) complete topology of the network. Examples of such schemes are the conventional routing schemes, Destination-Sequenced Distance Vector (DSDV). On the other hand, reactive protocols establish paths only upon request, for example, in response to a query, or an event; meanwhile, sensors remain idle in terms of routing behavior. Sensors forward each routing request to peers until it arrives at a sink; the latter will respond over the reverse communication path. Examples of reactive routing schemes are Ad hoc On-demand Distance Vector (AODV), Dynamic Source Routing (DSR). Hybrid protocols use a combination of these two ideas. Ayyaswamy Kathirvel and Rengaramanujam Srinivasan proposed a new protocol that modifies AODV to improve its performance.

In [5], Wang et al. has proposed a cross-layer joint routing and MAC-PHY design to achieve energy balance and energy efficiency simultaneously in WSN. The energy-balanced routing distributes the levels of residue energy evenly throughout the network, while the optimal transmission power control achieves further energy savings by adjusting the transmission power to meet the communication quality requirements at the receiver.

In [6], Chen and Terzis have presented a work for mobile sensor nodes based on Log-normal path loss model. They have started with the hypothesis that if the lognormal model holds perfectly, then the signal strengths at nearby locations are independent. In turn, this implies that finding the location whose packet reception ratio is above a certain threshold can be modeled as a sequence of Bernoulli trials. Therefore, the number of attempts required to find a location with high PRR is geometrically distributed. Furthermore, they argued that the geometric distribution is desirable because it does not exhibit long tails, and therefore the number of attempts to find a good location should usually be small, provided that a reasonable percentage of location with high PRR exists in the search vicinity. Rather than looking at signal strength variations in the time domain, they focused on variations in the space domain. Specifically, they studied the implications of the lognormal path loss model when deploying or moving sensor motes.

*2.1. Introduction to AODV.* Ad hoc On-Demand Distance Vector (AODV) Routing is a routing protocol for mobile ad hoc networks (MANETs) and other wireless ad hoc networks. It is jointly developed in Nokia Research Center, University of California, Santa Barbara and University of Cincinnati by C. Perkins, E. Belding-Royer, and S. Das. It is a reactive routing protocol, meaning that it establishes a route to a destination only on demand. It employs destination sequence numbers to identify the most recent path. When a node needs to determine a route to a destination node, it floods the network with a Route Request (RREQ) message.

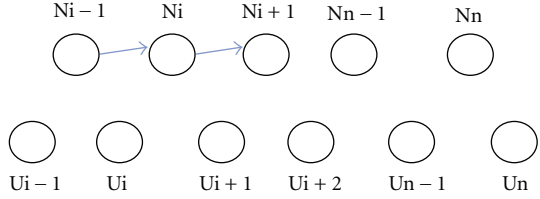

Figure 1: Triple umpiring system.

The originating node broadcasts a RREQ message to its neighboring nodes, which broadcast the message to their neighbors and so on. To prevent cycles, each node remembers recently forwarded route requests in a route request buffer. As these requests spread through the network, intermediate nodes store reverse routes back to the originating node. Since an intermediate node could have many reverse routes, it always picks the route with the smallest hop count. When a node receiving the request either knows of a "fresh enough" route to the destination or is itself the destination, the node generates a Route Reply (RREP) message and sends this message along the reverse path back towards the originating node. As the RREP message passes through intermediate nodes, these nodes update their routing tables, so that, in the future, messages can be routed though these nodes to the destination.

*2.2. Introduction to Triple Umpiring System.* In the umpiring system [7] as shown in Figure 1, each node is issued with a token at the inception. The token consists of two fields: Node ID and status. Node ID is assumed to be unique and deemed to be beyond manipulation; status is a single bit flag. Initially the status bit is preset to zero indicating a green flag. The token with green flag is a permit issued to each node, which confers it the freedom to participate in all network activities. Each node in order to participate in any network activity, say Route Request RREQ, has to announce its token. If status bit is "1" indicating "red flag," then the protocol does not allow the node to participate in any network activities. Gwalani et al. in [8] proposed a new protocol that modifies AODV to improve its performance. The protocol, AODV-PA, incorporates path accumulation during the route discovery process in AODV to attain extra routing information. They have shown from the results that AODV-PA improves the performance of AODV under conditions of high load and moderate-to-high mobility. Krco and Dupcinov in [9] observed a problem that affects the neighbor detection algorithm of the AODV-routing protocol and has a deteriorating impact on performance of ad hoc networks that use this protocol. An improvement of the neighbor detection algorithm [10] based on the differentiation of good and bad neighbors using signal to noise ratio (SNR) value is proposed, described, and experimentally verified.

## 3. MTUS Model

TUS [11, 12] can be modified to enable path accumulation during the route discovery cycle. When the Route Request (RREQ) and Route Reply (RREP) messages are generated or forwarded by the nodes in the network, each node appends its own address on these route discovery messages. Each node also updates its routing table with all the information contained in the control messages. As the RREQ messages are broadcast, each intermediate node that does not have a route to the destination forwards the RREQ packet after appending its address in the packet. Hence, at any point the RREQ packet contains a list of all the nodes traversed. Whenever a node receives a RREQ packet, it updates the route to the source node. It then checks for intermediate nodes accumulated in the path. Before making an entry, we propose differentiation between the so-called "good" neighbors and "bad" neighbors. Classification is done dynamically based on the signal to noise ratio (SNR) value that is measured whenever a packet that contains a TUS message is received. Neighbors are typically classified as "bad" if the quality of the interconnecting channel is poor; that is, it is not good enough to carry broadcast and unicast messages with sufficient quality regardless of transmission rate or coding technique.

## 4. MTUS with Optimal SNR-Based Power Control

In wireless signal transmission, one of the major sources of loss is attenuation. Basically the communication range decreases as the transmission data rate increases. One of the important parameter of interest is bit error rate (BER). The desirable BER value can be mapped into a desirable SNR value for a given modulation scheme. The desirable SNR value required by a given data rate increases with the data rate. That is, if data rate increases, the probability of error also increases, and a higher SNR value is required at the transmitter to achieve the same BER at the receiver. Hence power supply increases with the SNR value.

The relation between transmit power $P_S$ and the SNR value at the receiver ($\text{SNR}_{Rx}$) is given by

$$\text{SNR}_{Rx} = \frac{P_S}{N} \cdot A, \tag{1}$$

where "$A$" is the channel attenuation factor including antenna gain in transmission.

The noise power "$N$" can be expressed as

$$N = N_0 \cdot R_S, \tag{2}$$

where $N_0$ is the noise power density. The transmission symbol rate is given by

$$R_S = \frac{R}{b}, \tag{3}$$

where "$R$" is the transmission rate, and "$b$" is the modulation constellation size. Consider

$$\text{SNR}_{Rx} = \frac{E_b}{N_0} \cdot b, \tag{4}$$

where "$E_b$" is energy per bit

$$E_b = P_S * T_b. \tag{5}$$

The transmission time of each bit is as follows:

$$T_b = \frac{1}{R}. \qquad (6)$$

Hence the optimal transmission in terms of specific desirable BER at the receiver end can be expressed as

$$P_S = \text{SNR}_{Rx} \cdot N \cdot \frac{1}{A},$$
$$P_S = \text{SNR}_{Rx} \cdot R_S \cdot \frac{N_0}{A}, \qquad (7)$$
$$P_S = R_S \cdot b \frac{N_0}{A} \frac{E_b}{N_0}, \qquad (8)$$

where the factor "$A$" is the product of antenna gain and channel loss. We have

$$A = K * L^{-1}. \qquad (9)$$

The relationship between BER and SNR for quadrature phase shift keying (QPSK) is given by

$$\text{BER} = \frac{1}{2} \text{erfc} \sqrt{\frac{E_b}{N_0}}. \qquad (10)$$

Hence the ratio $E_b/N_0$ can be calculated from BER as follows:

$$\frac{E_b}{N_0} = \left[\text{erfc}^{-1}(2 \cdot \text{BER})\right]^2. \qquad (11)$$

From (8)

$$P_S = R_S \cdot b \cdot \left(\frac{N_0}{A}\right) \left[\text{erfc}^{-1}(2 \cdot \text{BER})\right]^2. \qquad (12)$$

The signal strength at the receiver can be calculated as follows.

In a receiver, this concept involves the following parameters.

*Minimum Detectable Signal Power (MDS).* It is dependent on the modulation type as well as the noise specs of the antenna and receiver.

*Maximum Allowable Signal Power (MAS).* It is limited by the compression or third-order intercept points.

*Minimum Detectable Signal (MDS).* For a given receiver noise power, MDS determines the minimum signal-to-noise ratio at the output of the receiver (SNRo). Typical minimum SNR for QPSK with $P_e = 10^{-5}$ is 10 dB. We have

$$S_{i\,\min} = KT_0 FB \left(\frac{S_0}{N_0}\right)_{\min}. \qquad (13)$$

In dB

$$S_{i\,\min}(\text{dBm}) = -174 + B_{(\text{dBHz})} + F(\text{dB}) + \left(\frac{S_0}{N_0}\right)_{\min \text{dB}}, \qquad (14)$$

where "$K$" is the Boltzmann constant, "$F$" is the receiver noise Figure, $B$ is the receiver bandwidth, and $T_0 = 290$ K.

The receiver dynamic range ($\text{DR}_r$) can be calculated as

$$\text{DR}_r = \frac{\text{MAS}}{\text{MDS}}. \qquad (15)$$

The measured receiver signal strength (aggregated value) can be fed back in the beacon message to let the transmitter to know the received signal strength. Based on the receiver feedback, the transmitter either increases or decreases the transmit power $P_S$ there by achieving optimal power reduction.

## 5. MTUS with Optimal Handoff-Based Self-Recovery Feature

We also did a modification in MTUS link breakage recovery [13] mechanism. In MTUS, the source node broadcasts RREQ message to find a new route to the destination when the link break is occurred. As an improvement of MTUS, self-recovery [10, 11] MTUS takes the intermediate node, which detects the link break, to repair the break route.

Once the intermediate node cannot repair the route in time, the backward prehop node tends to find a new route instead. MTUS tends to repair break route if the broken node is near to the destination node. Otherwise, if the break node is far away from destination node, a Route Error (RRER) message is sent back to source node, and the source node rebroadcasts RREQ to find a new route. Self-recovery MTUS can repair break route without considering the distance between the broken node and the destination node. Because the intermediate nodes are usually nearer than the source node to the destination, the intermediate nodes on the data flow are more suitable than the source to broadcast RREQ to repair or find a route to destination. In the optimal handoff-based self-recovery feature, the route maintenance is performed by detecting the link break before the complete failure of the link. For this purpose each node maintains a neighbors power list (NPL). The decision to predict a link break [14] is made on the basis of received power with which a node is received from its neighbors with whom it forms part of an active route. When a link break on a path of data delivery occurs, the intermediate node upstream of that break may choose to repair the link locally by itself [15] if its power level is greater than certain threshold. Otherwise the source will reinitiate a route discovery instead. The flow chart for optimal handoff-based self-recovery feature is shown in Figure 2. The proposed method will intimate the problem to the upper layers if the packet has not been received by the receiver after fifteen trials.

## 6. Experimental Results and Analysis

We use a simulation model based on Glomosim-2.03 [16] in our evaluation.

Our performance evaluations are based on the simulations of 500 wireless sensor nodes that form a wireless sensor network over a rectangular (1000 × 1000 m) flat space.

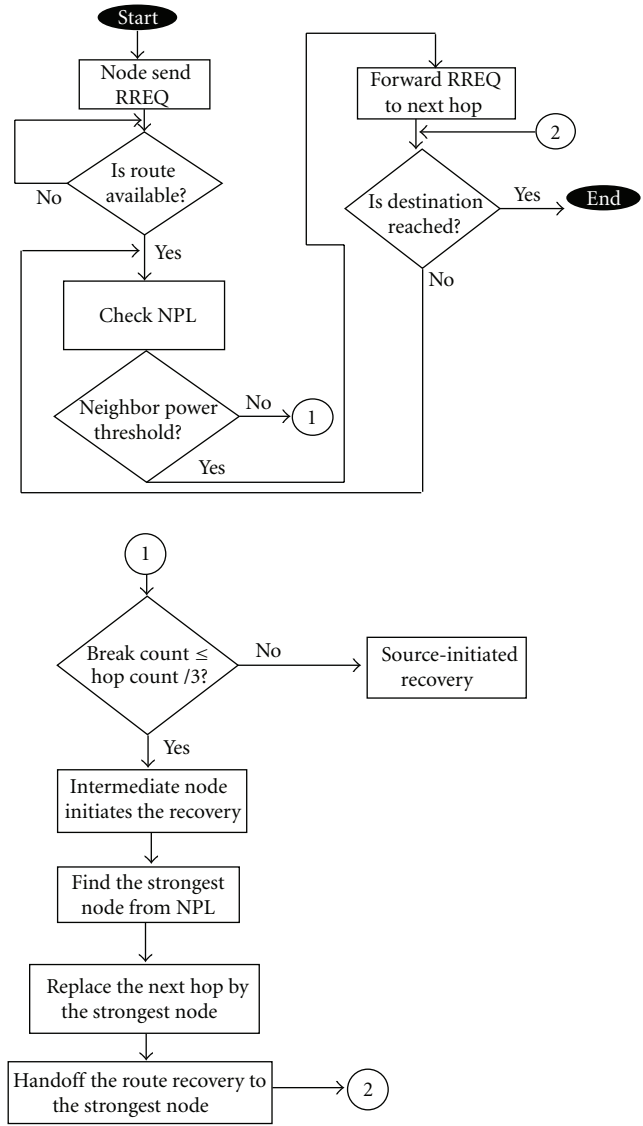

Figure 2: Flow chart for optimal hand off based self recovery.

Table 1: Simulation parameters.

| | |
|---|---|
| Area of sensing field | 1000 ∗ 1000 m |
| Number of sensor nodes | 500 |
| Simulation time | 600 s |
| Frequency | 2.4 GHz |
| Bandwidth | 2 Mbps |
| Traffic type | Constant bit rate (CBR) |
| Payload size | 30 to 70 Bytes |
| Number of loads | 200 Packets |
| Number of nodes | 500 nodes |
| Propagation limit (dbm) | −111.0 |
| Path loss model | Two ray model |

The MAC layer protocol used in the simulations was the distributed coordination function (DCF) of IEEE 802.11. The performance setting parameters are given in Table 1. We fix the distance between the source and sink to be 350 meters. The other 498 nodes are deployed between the source-sink pair. We assume that all nodes are able to adjust transmission power arbitrarily. Thus all nodes can send packets directly to any other node and allocate optimal transmission power for all possible link state information. We assume that channel conditions remain fixed during the time period of the end-to-end transmission of every packet. Each flow did not change its source and destination for the lifetime of a simulation run. We also evaluate the performance of all routing strategies under two different transmission power settings:

(1) Tx power: 15 dBm (Transmission range is 376.782 m).
(2) Tx power: 10 dBm (Transmission range is 282.547 m).

*6.1. Setting Up the Transmission Range.* The radio range is the average maximum distance in usual operating conditions between two nodes. There is no standard and common operating procedure to measure a range (except in free space, which is useless), so we cannot really compare different products from the ranges as indicated in the mobile devices data sheets. If we want to compare mobile nodes in terms of range performance, we must look closely at the *transmitted power* and *sensitivity values*. These are some measurable characteristics of the hardware which indicate the performance of the products in that respect. The transmitted power is the strength of the emissions measured in Watts (or milliWatts). Government regulations limit this power, but also having a high transmit power will also be likely to drain the batteries faster. Nevertheless, having a high transmit power will help to emit signals stronger than the interferers in the band. The sensitivity is the measure of the weakest signal that may be reliably heard on the channel by the receiver (it is able to read the bits from the antenna with a low error probability). This indicates the performance of the receiver, and the lower the value the better the hardware. Usual values are around −80 dBm (the lowest, the better, e.g., −90 dBm is better). A possible methodology to determine the transmission radio range in GloMoSim would be the following.

(1) Set the propagation pathloss model (PROPAGATION-PATHLOSS parameter).
(2) Fix the received power of the destination antenna (RADIO-RX-THRESHOLD parameter).
(3) Fix the distance and calculate the transmitted power according to the selected propagation pathloss model.
(4) Set this value to the RADIO-TX-POWER parameter.

GloMoSim has the following propagation models: free space and the ground reflection (or two-ray) models. The free space propagation model is used to predict received signal strength when the transmitter and receiver have a clear, unobstructed line-of-sight between them. This model predicts that transmission power is attenuated in proportion to the square of the distance. The ground reflection (two-ray) model considers both the direct path and a ground-reflected propagation path between transmitter and receiver. This model predicts that received power falls of distance raised to the fourth power, or at a rate of 40 dB/decade. This is a much more rapid path loss than is experienced in free space. We have

$$P_r = P_t \frac{h_t^2 h_r^2}{d^4} G_t G_r, \quad (16)$$

where $h_t$ and $h_r$ for the height of the transmitter and receiver antennas, and in GloMoSim this value is hard coded to 1.5 meters. The antenna gain (represented in GloMoSim by the RADIO-ANTENNA-GAIN parameter) is a measure of the directionality of an antenna. Antenna gain is defined as the power output, in a particular direction, compared to that produced in any direction by a perfect omnidirectional antenna (isotropic antenna). In both models, we have to measure the loss or attenuation of signal strength. The *decibel* is a measure of the ratio between two signal levels. The decibel gain is given by the following equation:

$$G_{\text{dB}} = 10 \log_{10} \frac{P_{\text{out}}}{P_{\text{in}}}. \quad (17)$$

It is convenient to be able to refer to an absolute level of power in decibels so that gains and losses with reference to an initial signal level may be calculated easily. The dBW (*decibel Watt*) is used extensively in microwave applications. The value of 1 W is selected as a reference and defined to be 0 dBW. Another common unit is the dBm (*decibel milliWatt*), which uses 1 mW as the reference. Thus 0 dbM = 1 mW. This is represented in the following formulae:

$$\text{Power}_{\text{dbW}} = 10 \log \frac{\text{Power}_W}{1\,W}, \quad \text{Power}_W = 10^{\text{Power}_{\text{dbW}}/10},$$

$$\text{Power}_{\text{dbm}} = 10 \log \frac{\text{Power}_{\text{mW}}}{1\,\text{mW}}, \quad \text{Power}_{\text{mW}} = 10^{\text{Power}_{\text{dbW}}/10}, \quad (18)$$

PROPAGATION-PATHLOSS = TWO-RAY, PROPAGATION-LIMIT = −111 (dBm), RADIO-FREQUENCY = 2.4 e 9 (hertz), RADIO-TX-POWER = 15 (dBm), RADIO-RX-THRESHOLD = −81 (dBm), RADIO-ANTENNA-GAIN = 0.0 (dBm).

The propagation pathloss model indicates us to use the two-ray model formula as follows:

$$G_t = G_r = 0\,\text{dBm}, \quad (19)$$

$$P_r = P_t \frac{h_t^2 h_r^2}{d^4} G_t G_r,$$

$$d = \sqrt[4]{\frac{P_t h_t^2 h_r^2}{P_r}} = \sqrt[4]{\frac{(31.6227\,\text{mW})(1.5)^2(1.5)^2}{7 \times 10^{-9}}} = 388\,\text{m}. \quad (20)$$

We compared our ESRP with two different transmission power levels based on the following output parameters.

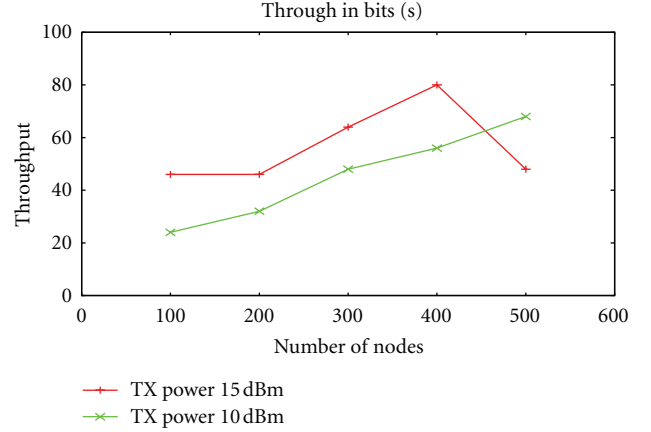

Figure 3: Number of nodes versus throughput.

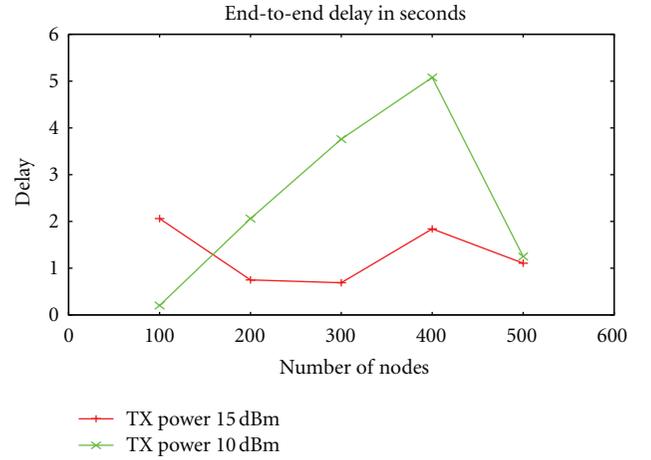

Figure 4: Number of nodes versus end to end delay.

*(1) Packet delivery ratio.* It is the ratio of the number of data packets successfully delivered to the destinations to those generated by the sources. Consider

$$\text{PDR} = \frac{N_r}{N_t}, \quad (21)$$

where $N_r$ is the number of data packets successfully received, and $N_t$ is the number of data packets transmitted.

*(2) End to end delay (Seconds).* It indicates the time taken for the message to reach from source to destination.

*(3) Energy consumption in mWH.* Figure 3 shows that the reduction in the TX power helps to improve the throughput as the number of nodes increases. Figure 4 shows that the end-to-end delay considerably decreases, with lower transmission power, even though it increases initially. The average energy consumption is reduced from 40 mWH to 30 mWH when TX power is reduced from 15 dBm to 10 dBm as shown in Figure 5.

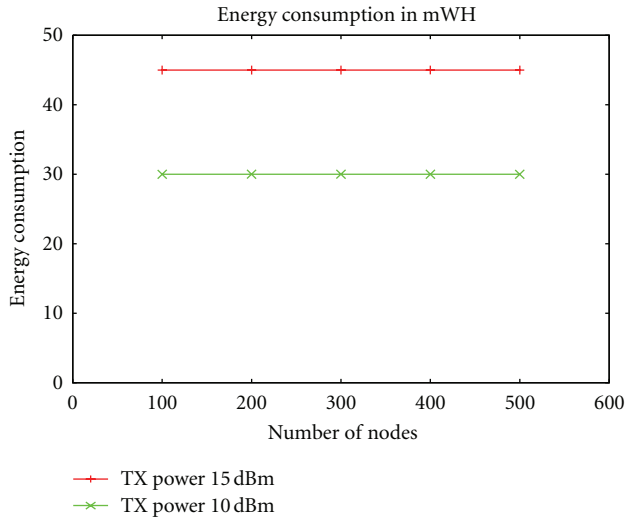

Figure 5: Number of nodes versus energy consumption.

## 7. Conclusion

We are attempting to develop a comprehensive approach to understand the fundamental performance of information routing in energy-limited wireless sensor networks through optimal SNR- [17, 18] based power control mechanism and optimal handoff-based self-recovery features. We presented some results for a few different small-scale WSN experiments to study the solutions obtained for these problems as we vary the fairness constraints [19, 20]. We found that higher fairness constraints can result in significant decrease in information extraction and higher energy usage. Another observation about the results is that the flow and energy curves show qualitatively abrupt changes as the fairness constraints are varied. Based on the simulation results, we can conclude that efficient and secure routing protocol (ESRP) with optimal power control mechanism and handoff-based self-recovery can significantly reduce the power usage. We note that this is a very much work in progress. We are currently trying to make the models richer and more useful for analyzing different kinds of wireless sensor networks. One significant extension [21] would be to incorporate in- network aggregation to capture the data-centric nature of these systems. Other extensions we are looking into include in-depth analysis of the impact of other parameters [22] such as sensor deployment/placement, different modulation and path loss methods, and energy and information extraction constraints. In the longer term, we also hope to enhance the optimization-based formulations with closed-form analytical expressions [23].

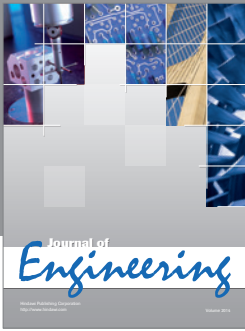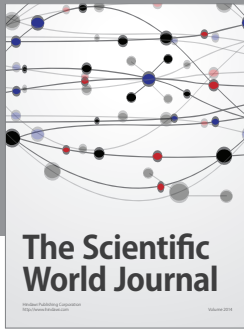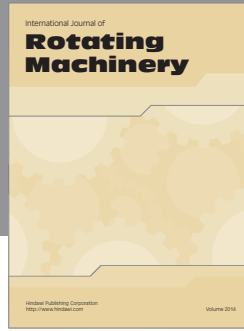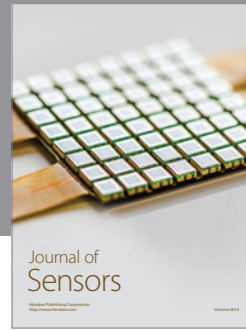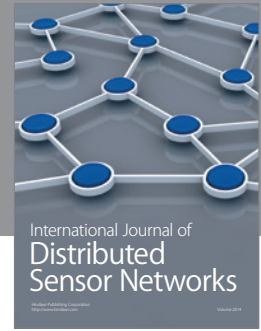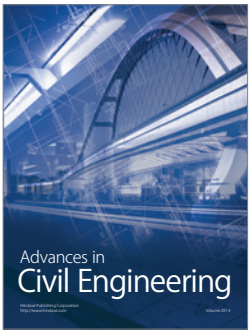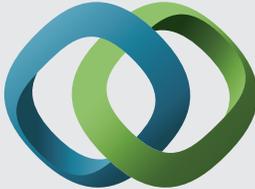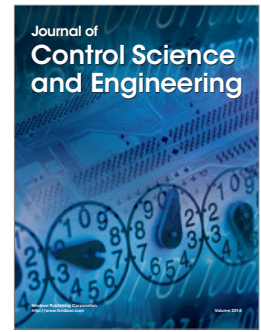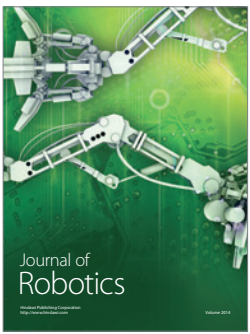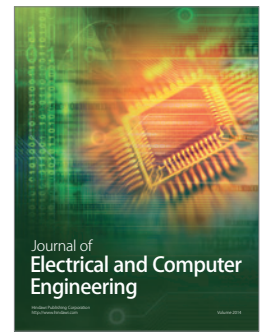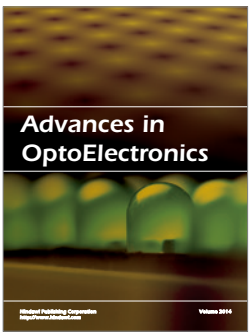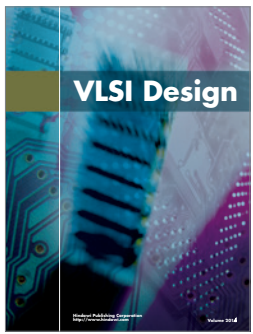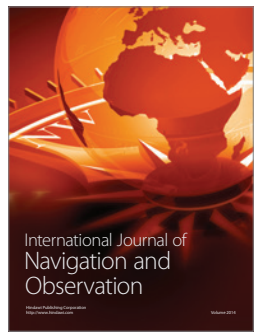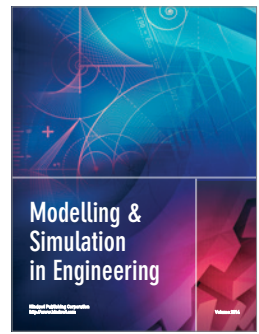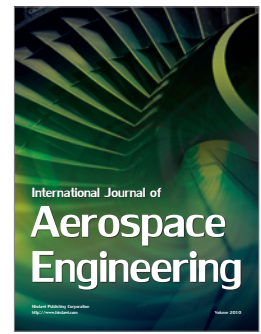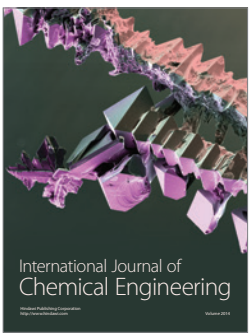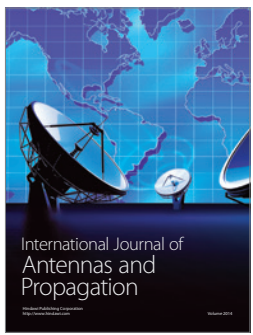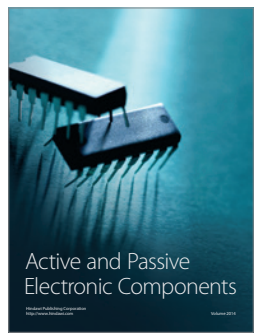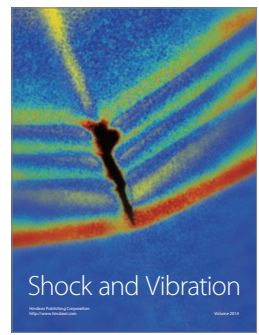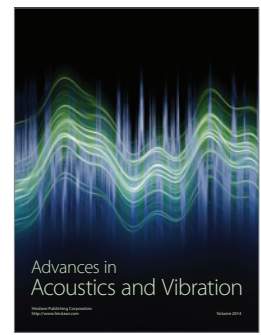